\documentstyle[prd,aps]{revtex}
\input psfig.sty
\input epsfig.sty
\tightenlines
\newcommand{\be}{\begin{equation}}
\newcommand{\ee}{\end{equation}}
\newcommand{\bea}{\begin{eqnarray}}
\newcommand{\beas}{\begin{eqnarray*}}
\newcommand{\eea}{\end{eqnarray}}
\newcommand{\eeas}{\end{eqnarray*}}
\newcommand{\ba}{\begin{array}}
\newcommand{\ea}{\end{array}}
\begin{document}
\draft

\title{ Ward identity in   noncommutative QED}

\author{ T. Mariz~\footnote{E-mail address: 
tiago@fisica.ufpb.br}, C. A. de S. Pires~\footnote{E-mail:cpires@fisica.ufpb.br}, and
R. F. Ribeiro~\footnote{E-mail address: rfreire@fisica.ufpb.br} 
}
\address{ \it  Departamento de F\'{\i}sica, Universidade Federal da Para\'{\i}ba, 
Caixa Postal 5008, 58051-970, Jo\~ao
Pessoa PB, Brazil.   }
\date{\today}
\maketitle
\vspace{.5cm}

\hfuzz=25pt
\begin{abstract}
 Although noncommutative QED presents a  nonabelian structure, it does not present  structure constants.  In view of this we investigate  how Ward identity is satisfied in  pair annihilation process and  $\gamma \gamma \rightarrow \gamma \gamma$ scattering in noncommutative QED.  
\end{abstract}
\vskip2pc
\newpage
\section{introduction}

In  1947 Snyder perceived  that the introduction of a smallest unit of length in space-time  forces the drop of
 the usual assumption of commutativity among the space-time coordinates\cite{snyder}.  Behind 
such idea was the attempt of solving divergence in matter-field interactions. The idea received few attention until  1999 when new  developments  in string theory revealed that noncommutative 
space-time is realized in string theory  when open 
string propagates in the presence of constant background antisymmetric tensor field\cite{connes,s-w}. Motivated by such
 theoretical achievement the idea was soon extended to  quantum field theory\cite{douglas}. In noncommutative quantum
 field theory(NCQFT) much attention has been devoted to noncommutative QED(NCQED)\cite{scalar,riad,hayakama,susskind,lorentz,unitarity,frenkel,yang-mills,varios,review}. The reason for this is found in  the structure and in the questions raised in the  NCQED. Even though NCQED is a gauge theory based on the  symmetry  $U(1)_{em}$, its structure   presents a nonabelian character\cite{douglas}. As a direct consequence of the noncommutativity among space-time coordinates, the theory is no longer Lorentz invariant\cite{lorentz} neither respect
 unitatity if time does not commute with space\cite{unitarity}. Other novelty is that there appears a connection among ultraviolet and infrared divergences\cite{riad,hayakama,susskind,lorentz,unitarity}.

From the phenomenological point of view,  much  attention has been given  to the  nonabelian character of NCQED\cite{mathews,chair,gamma}. It is well know that the unique interaction in ordinary QED is the current interaction. In NCQED  this interaction  changes by a  momentum-dependent phase factor. Besides, it now disposes of 3 and 4 point interactions 
giving rise to the  $\gamma \gamma \rightarrow \gamma \gamma$ scattering at tree level\cite{gamma}. Due to these changes, a re-analysis of all the basic processes in NCQED  appears to be necessary\cite{mathews,gamma}. In this regard, particular attention has been given to the processes  $e^+ e^- \rightarrow \gamma \gamma$,  and  $\gamma \gamma \rightarrow \gamma \gamma$\cite{gamma}. The reason for this  is that the first process receives a new contribution and the other exist only  due to the nonabelian character of NCQED. 

In the works\cite{mathews,gamma} attention has been called to the validity of the Ward identity. Differently from ordinary QED, where Ward identity is straightforward, in NCQED some care has to be taken due to the nonabelian structure of
the theory.
 We think this is  sufficient reason for a detailed analysis, and for an explicit checking of the Ward identity for the $e^+ e^- \rightarrow \gamma \gamma$  and  $\gamma \gamma \rightarrow \gamma \gamma$ processes.

The aim of this work is to check how the Ward identity comes about at tree level in NCQED in both processes  $e^+ e^- \rightarrow \gamma \gamma$  and 
$\gamma \gamma \rightarrow \gamma \gamma$\cite{comment}. For pedagogical reasons we make analogy with QCD. 
In QCD the structure constants  play  crucial role in getting the Ward identity  in  $gg \rightarrow gg$ 
scattering. It is the Jacobi identity among the structure
 constants of the $SU(3)_C$ group that assures the Ward identity. Although NCQED presents a nonabelian structure it does not present a  group structure, hence  there are no structure constants. There in place of the structure constants we have momentum-dependent phase factors.  Our  main question here is if  those phase factors will
 play the role of the structure constants through some equivalent Jacobi identity. 

To achieve these goals, we organize this work  in the following way. In Sec.~\ref{sec2} we begin discussing  about  the main aspects of NCQED, and after  in Sec.~\ref{sec3} we check the vality of  Ward identity  in the pair annihilation process. In Sec.~\ref{sec4} we check the identity in the $\gamma \gamma \rightarrow \gamma \gamma$ scattering. We end  our work in Sec.~\ref{sec5} where we introduce some comments and remarks.

\section{ noncommutative QED}
\label{sec2}
The idea behind  noncommutative  space-time is that in some very microscopic regime our common understanding of  
 space-time is not applicable anymore. Such regime is marked by a patch of area $\theta$ where space-time loses its condition of continuum and passes to obey the relation
\be
[\hat{x}_\mu ,\hat{ x}_\nu ] = i\theta_{\mu \nu},
\label{ncr}
\ee
where   $\theta_{\mu \nu}$ is a real antisymmetric  constant matrix. In the original idea $\theta_{\mu \nu}$ was an operator and then Lorentz invariance was preserved. Here $\theta_{\mu \nu}$ is an ordinary area. This  gives a preferential direction to space-time thus leading to violation of  Lorentz invariance.

One way of implementing  noncommutative coordinates in the  context of field theory is through the Moyal product\cite{connes}  
\be
A(x) \star  B(x) \equiv [ e^{(i/2)\theta_{\mu \nu}\partial_{\zeta\mu} \partial_{\eta \nu}}A(x+\zeta)B(x+\eta)]_{
\zeta=\eta=0}.
\label{operatorproduct}
\ee
The procedure with noncommutative coordinates goes like this. First the  Lagrangian is formulated in  terms of star $\star$
product. After we must change the   $\star$ product by the Moyal expansion in (\ref{operatorproduct}) in order to leave
 the Lagrangian in terms of  ordinary product. 

In gauge theory first thing to do is to express the gauge transformation in terms of $\star$ products
\be
A_\mu \rightarrow U\star A_\mu  \star U^{-1} + \frac{i}{g}U\star \partial_\mu U^{-1}.
\label{gaugeinv1}
\ee
In the case of NCQED the gauge symmetry is  $U=e^{i\alpha q}$.  With this symmetry the gauge invariance of the  photon takes the form  
\be
A_\mu \rightarrow A_\mu + \partial_\mu \alpha +2\sin(p_1 \theta p_2/2)A_\mu \alpha.
\label{gauginv2}
\ee
Perceive that such transformation is similar to a  nonabelian one. As immediate consequence the tensor $F^{\mu \nu}$ must change in order to the action of the  NCQED preserves the gauge invariance
\be
F_{\mu \nu}=\partial_\mu A_\nu -\partial_\nu A_\mu-ig[A_\mu,A_\nu]_\star=\partial_\mu A_\nu -\partial_\nu A_\mu-ig( A_\mu \star A_\nu -A_\nu \star A_\mu).
\label{tensor}
\ee
Applying the Moyal product, the tensor above takes the form
\be
F_{\mu \nu}= \partial_\mu A_\nu -\partial_\nu A_\mu +2g\sin(p_1\theta p_2/2)A_\mu A_\nu ,
\label{yangstruct}
\ee
and this leads us to conclude  that the  nonabelian character of NCQED is a pure geometric effect.

\begin{figure}[htbp]
\centerline{
\psfig{figure=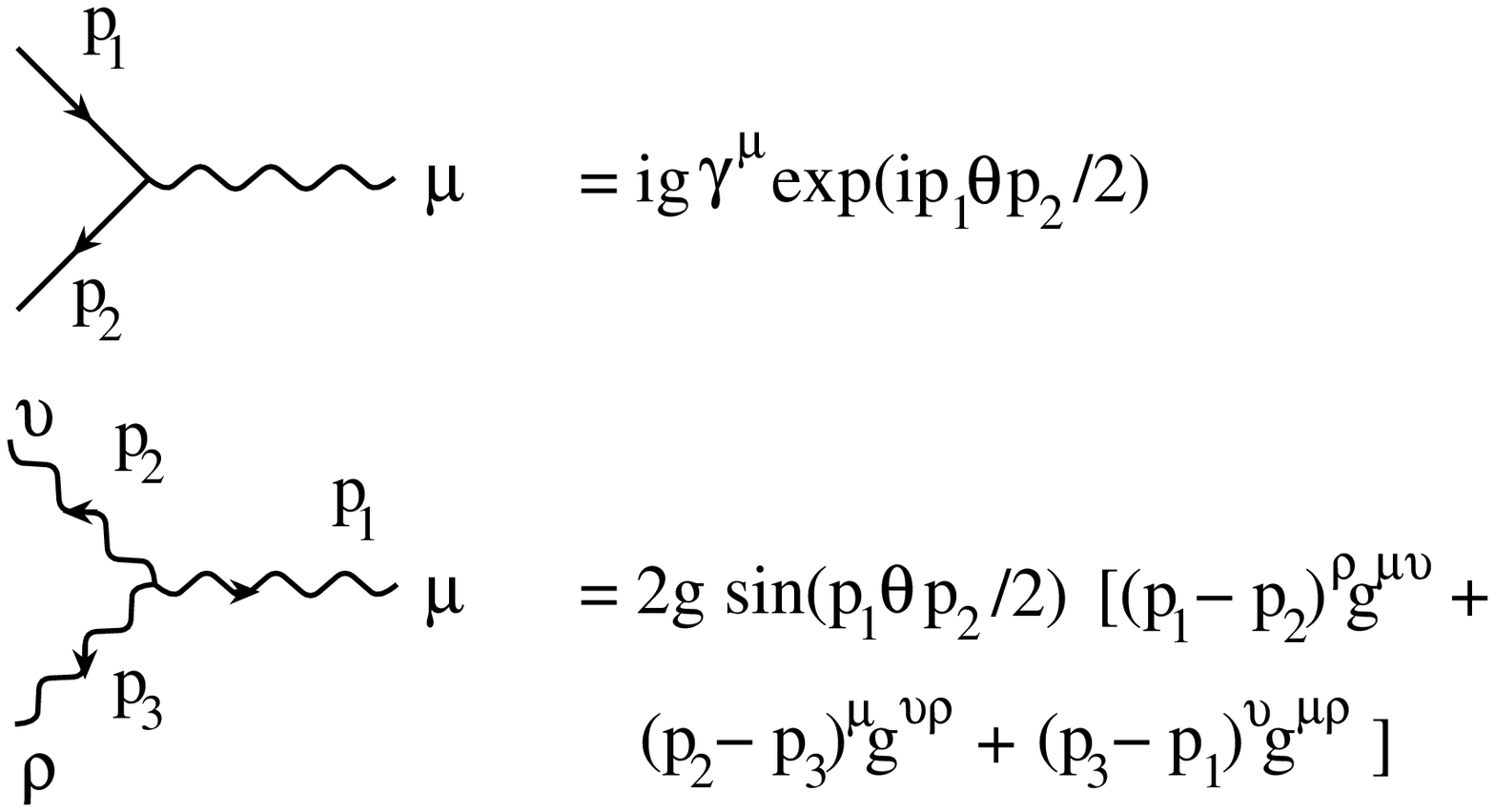,height=8.0cm,width=10.0cm,angle=0}}
\vspace*{-2.3cm}
\centerline{
\psfig{figure=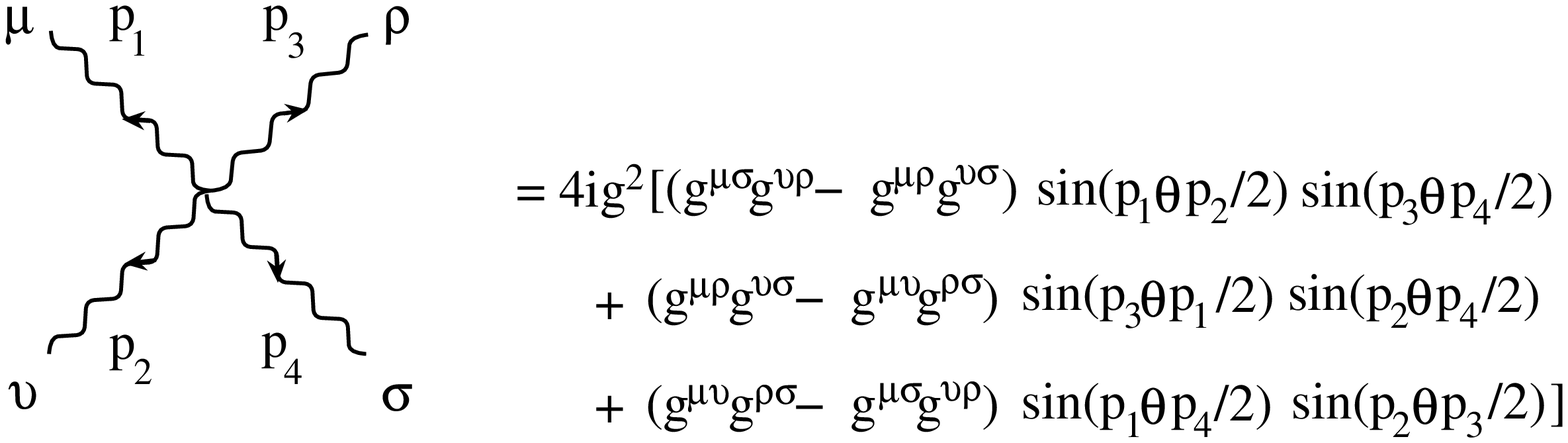,height=8.0cm,width=10.0cm,angle=0}}
\vspace*{-4.5cm}
\caption{Feynman rules of NCQED.}
\label{frules}
\end{figure}

 In terms of ordinary product, the NCQED presents the following action
\be
S = \int d^4x \left(- \frac{1}{4\pi}
F^{\mu \nu}F_{\mu \nu}+\bar \psi i\partial \!\!\!/ \psi -ge^{ip_1\theta p_2/2}\bar \psi A\!\!\!/ \psi -m \bar \psi \psi  \right).
\label{action}
 \ee
The Feynman rules drew from this action are displayed in FIG. (\ref{frules}).

\noindent

\section{Ward identity in pair annihilation process}
\label{sec3}
\begin{figure}[htbp]
\centerline{
\psfig{figure=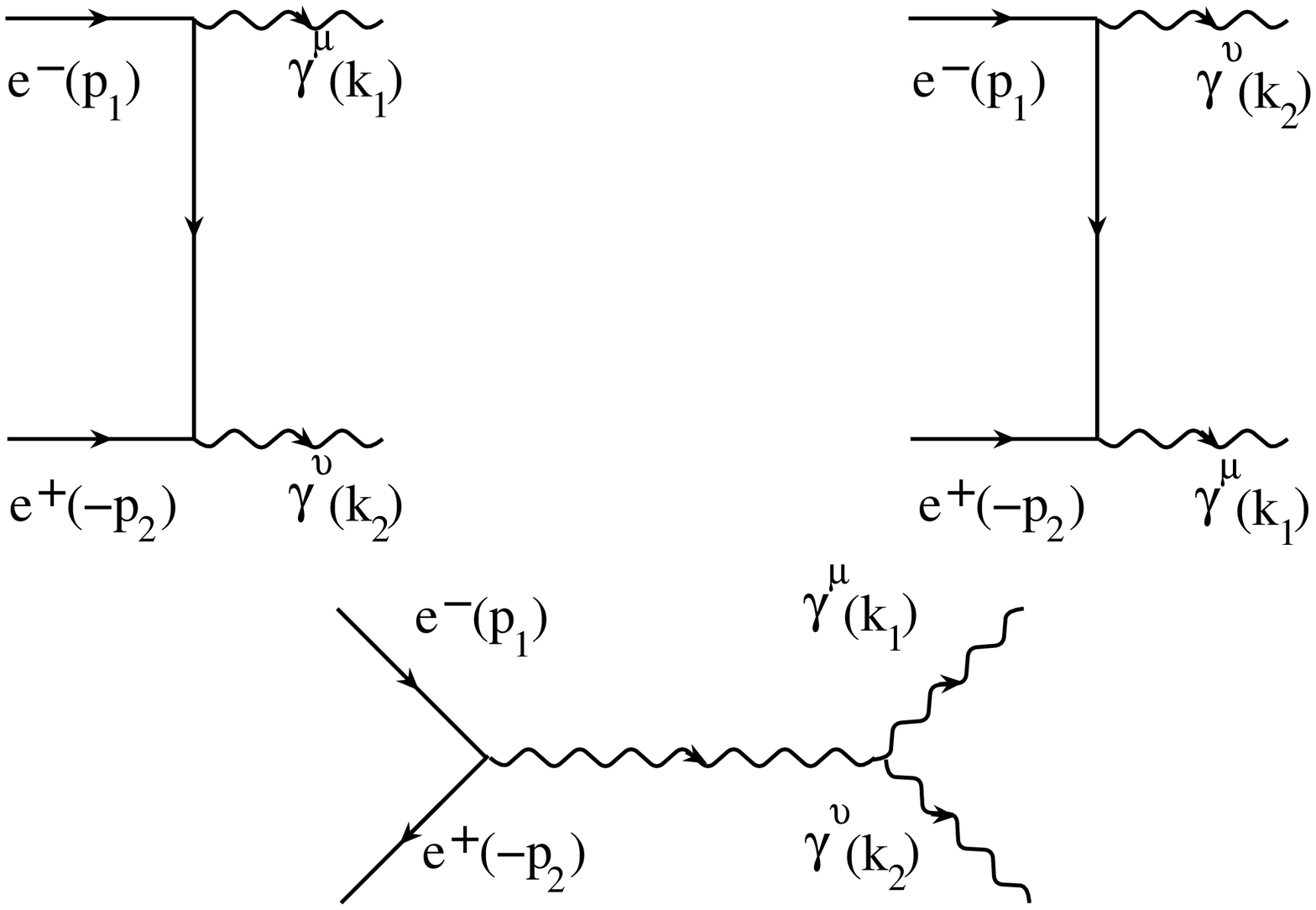,height=8.0cm,width=10cm,angle=0}}
\vspace*{0.5cm}
\caption{The three tree level contributions to $e^+ e^- \rightarrow \gamma 
\gamma$ in NCQED.}
\label{feynpair}
\end{figure}

 In NCQED the  $ e^+ e^- \rightarrow \gamma \gamma $  process gains a new contribution due to the 3 point interaction among the photons displayed in FIG. (\ref{feynpair}). The central issue  of this section is to see how Ward identity is satisfied in this process in view of the new contribution.

\noindent

 The total amplitude for the  $ e^+ e^- \rightarrow \gamma \gamma $ process contains three contributions
\be
i{\cal M}^{\mu \nu} \epsilon^*_\mu(k_1) \epsilon^*_\nu(k_2)=
i{\cal M}^{\mu \nu}_1 \epsilon^*_\mu(k_1) \epsilon^*_\nu(k_2) +i{\cal M}^{\mu \nu}_2 \epsilon^*_\mu(k_1)
\epsilon^*_\nu(k_2) + i{\cal M}^{\mu \nu}_3 \epsilon^*_\mu(k_1) \epsilon^*_\nu(k_2).
\label{amplitude1}
\ee
 According to the Feynman rules in FIG. (\ref{frules}), the first two  invariant amplitudes take the following form 
\bea
&&i{\cal M}^{\mu \nu}_{1,2} \epsilon^*_\mu(k_1) \epsilon^*_\nu(k_2)=
i{\cal M}^{\mu \nu}_1 \epsilon^*_\mu(k_1) \epsilon^*_\nu(k_2) +i{\cal M}^{\mu \nu}_2 \epsilon^*_\mu(k_1)
\epsilon^*_\nu(k_2)= \nonumber \\
&&(ig)^2 \bar v(p_2)\gamma^\mu e^{ip_1\theta (p_2-k_2)/2} \frac{i}{p_2\!\!\!\!\!/ -k_2\!\!\!\!\!/ -m}\gamma^\nu 
e^{ip_2 \theta (p_2-k_2)/2} u(p_1)\epsilon^*_\mu (k_1) \epsilon^*_\nu (k_2) \nonumber \\
&& +(ig)^2  \bar v(p_2)\gamma^\nu e^{ ip_2 \theta (k_2-p_1)/2 } \frac{ i }{ k_2\!\!\!\!\!/ -p_1\!\!\!\!\!/ -m }\gamma^\mu e^{ip_1
\theta (k_2-p_1)/2}u(p_1)\epsilon^*_\mu (k_1) \epsilon^*_\nu (k_2).
\label{amp1}
\eea

 Differently from QED we need some assumptions for the validity of the Ward identity. In choosing the photon of momentum $k_2$ we ought to  assume that the other is on shell, $k^2_1 =0$, and   transverse, $\epsilon \cdot k_1 =0$. 

In replacing 
$ \epsilon^*_\nu (k_2)$ by $ k_{2\nu}$, eliminating $k_2$ by momentum conservation, and also making use  of  the Dirac equations $(p_1\!\!\!\!\!/ -m)u(p_1)=0$
 , $\bar v(p_2)(-p_2\!\!\!\!\!/ -m)=0$, we are able to bring  the contraction of the amplitude with the momentum $k_2$  to the following simple form
\be
i{\cal M}^{\mu \nu}_{1,2} \epsilon^*_\mu(k_1) k_{2\nu} =2g^2 e^{ip_1\theta p_2/2} \sin[(p_1 + p_2)\theta k_2/2]\bar
 v(p_2)\gamma^\mu u(p_1) \epsilon^*(k_1)_\mu.
\label{amp2}
\ee 

Let us now  work out the third contribution. The amplitude of the third graphic in Fig. (\ref{feynpair}) is 
\bea
i{\cal M}^{\mu \nu}_3 \epsilon^*_\mu(k_1) \epsilon^*_\nu(k_2)&&=ig \bar v(p_2) \gamma_\rho 
e^{ip_1 \theta p_2/2}u(p_1)\frac{-i}{k^2_3} \epsilon^*_\mu(k_1) \epsilon^*_\nu(k_2) \nonumber \\
&& \times2g\sin(k_2\theta k_1/2)[g^{\mu \nu}(k_2 - k_1)^\rho + g^{\nu \rho} (k_3 -k_2)^\mu + g^{\rho \mu}(k_1 - k_3)].
\label{amp3}
\eea
In replacing $\epsilon^*_\nu(k_2)$ by $ k_{2\nu}$ and   using the momentum conservation, $p_1 +p_2 = k_1 +k_2$,   we obtain after some manipulation
\be
i{\cal M}^{\mu \nu}_3 \epsilon^*_\mu(k_1) k_{2_\nu}=-2g^2e^{ip_1 \theta p_2/2}\sin[(p_1 + p_2)\theta k_2/2)]
 \bar v(p_2) \gamma^\mu u(p_1) \epsilon^*_\mu(k_1).
\label{amp4}
\ee
From  (\ref{amp2})  and (\ref{amp4}) we have that the Ward identity is satisfied
\be
i{\cal M}^{\mu \nu} \epsilon^*_\mu(k_1) k_{2\nu}=0.
\label{check1}
\ee 
 We think that  this check is necessary once it is not trivial that
 the structure of the amplitudes in  (\ref{amp1}) and (\ref{amp3})  leads to a cancellation when summed. It is interesting to see that the cancellation occurs without resort to any suppositions over the momentum-dependent  phase factors.

\section{ Ward identity in  $ \gamma \gamma \rightarrow \gamma\gamma$ scattering}
\label{sec4}
It is very well known that  photons do not carry any kind of charge. Then they do not present self interactions.
  This is no longer true in scenarios involving noncommutative space-time. In  NCQED  photons present 3 and 4 point 
interactions. This gives rise to  $ \gamma \gamma \rightarrow \gamma\gamma$ scattering at tree level. The analysis of
 such scattering has been carried out  by many authors\cite{gamma}. However the checking of the Ward identity of such 
scattering was not done yet. In those works  the Ward identity is assumed to be  valid. In view of this, it turns 
useful to check  the Ward identity in  $ \gamma \gamma \rightarrow \gamma\gamma$ scattering in  NCQED. Also we find interesting  to make some analogy with QCD since there the structure constants play an important role, through the Jacobi identity, in getting  the  Ward 
identity in  $gg \rightarrow gg$ scattering.

\noindent
\begin{figure}[htbp]
\centerline{
\psfig{figure=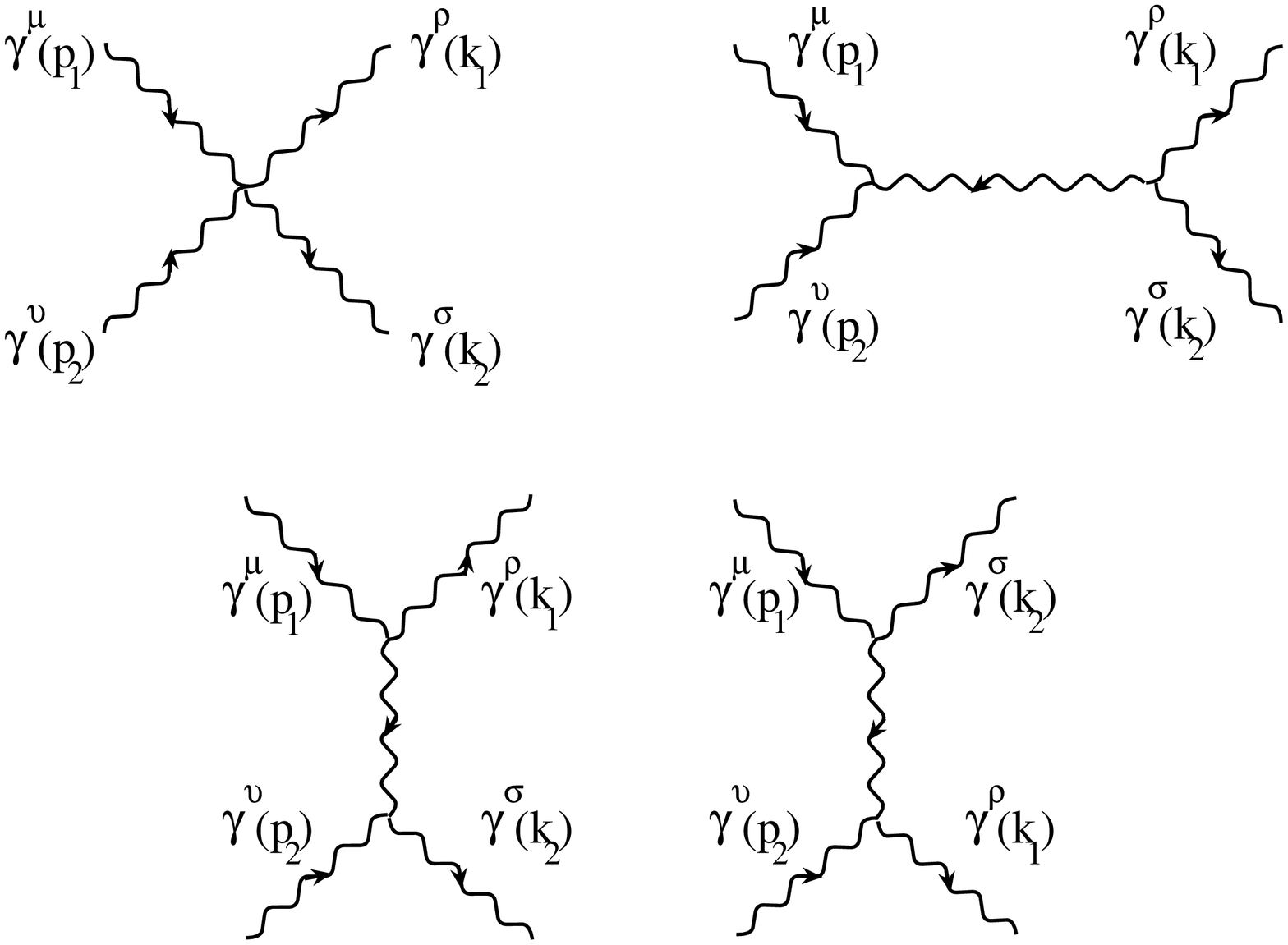,height=8.0cm,width=10cm,angle=0}}
\vspace*{0.5cm}
\caption{The tree level contributions to $\gamma \gamma \rightarrow \gamma 
\gamma$ in NCQED.}
\label{fgamma}
\end{figure}

The four contributions for the $ \gamma \gamma \rightarrow \gamma\gamma$ scattering is presented in FIG. (\ref{fgamma}).  The first
 three graphics compose the channels $s$, $t$ and $u$. The respective amplitudes are 
\bea
i{\cal M}^{\mu \nu \rho \sigma}_s \epsilon_\mu(p_1) \epsilon_\nu(p_2) \epsilon_\rho^*(k_1) \epsilon_\sigma^*(k_2)&=& -4ig^2\sin(p_1 \theta p_2/2)\sin(k_1 \theta k_2/2)\epsilon_\mu(p_1) \epsilon_\nu(p_2) \epsilon_\rho^*(k_1) \epsilon_\sigma^*(k_2)\nonumber
 \\ &&\times C^{\alpha \mu \nu}(-p_1,-p_2,q_1) \frac{g_{\alpha \beta}}{q_1^2}C^{\beta \rho \sigma}(-k_1,-k_2,q_1),\nonumber \\
&&\nonumber \\
i{\cal M}^{\mu \nu \rho \sigma}_t \epsilon_\mu(p_1) \epsilon_\nu(p_2) \epsilon_\rho^*(k_1) \epsilon_\sigma^*(k_2) &=& -ig^2\sin(p_1 \theta k_1/2)\sin(p_2 \theta k_2/2)\epsilon_\mu(p_1) \epsilon_\nu(p_2) \epsilon_\rho^*(k_1) \epsilon_\sigma^*(k_2)\nonumber \\
&&\times C^{\mu \alpha \rho}(q_2,k_1,-p_1)\frac{g_{\alpha \beta}}{q^2}C^{\sigma \nu \beta}(p_2,q_2,-k_2),\nonumber \\
&&\nonumber \\
i{\cal M}^{\mu \nu \rho \sigma}_u \epsilon_\mu(p_1) \epsilon_\nu(p_2) \epsilon_\rho^*(k_1) \epsilon_\sigma^*(k_2)&=& -ig^2\sin(p_1 \theta k_2/2)\sin(p_2 \theta k_1/2)\epsilon_\mu(p_1) \epsilon_\nu(p_2) \epsilon_\rho^*(k_1) \epsilon_\sigma^*(k_2)\nonumber \\
&&\times C^{\mu \alpha \sigma}(q_3,k_2,-p_1)\frac{g_{\alpha \beta}}{q^2_3}C^{\rho \nu \beta}(p_2,q_3,-k_1),
\label{ampqcdstu1}
\eea
where
\be
C^{\theta \phi \gamma}(P_1,P_2,P_3)= (P_1 - P_2)^\theta g^{\phi \gamma}+(P_2 - P_3)^\phi g^{\gamma \theta} + (P_3 -P_1)^\gamma g^{\theta \phi}.
\label{C}
\ee
To go further we  choose one photon for the checking. Let us take the photon of momentum $k_2$,  and then suppose that all the other three are transverses,  $\epsilon(k_1) \cdot k_1 =\epsilon(p_1)\cdot p_1 = \epsilon(p_2)\cdot p_2=0$ and on shell,  $p_1^2 =0$, $p_2^2 =0$,  $k_1^2 =0$. After this we replace
 $\epsilon_\sigma(k_2)$ by $ k_{2 \sigma}$ in the above amplitudes.  First thing to note here is that 
\bea
&& \epsilon_\rho^*(k_1)k_{2\sigma} C^{\beta \rho \sigma}(-k_1,-k_2,q_1) =\epsilon_\rho^*(k_1)(q^2_1g^{\rho \beta} -q_1^\beta q_1^\rho),\nonumber \\
&&\nonumber \\
&&\epsilon_\nu(p_2)k_{2\sigma}C^{\sigma \nu \beta}(p_2,q_2,-k_2)=-\epsilon_\nu(p_2)(q^2_2g^{\nu \beta} -q_2^\beta q_2^\nu),\nonumber \\
&&\nonumber \\
&&\epsilon_\mu(p_1)k_{2\sigma}C^{\mu \alpha \sigma}(q_3,k_2,-p_1)=\epsilon_\mu(p_1)(q^3_2g^{\mu \alpha} -q_3^\mu q_3^\alpha).
\label{rel2}
\eea

Substituting (\ref{rel2})  in (\ref{ampqcdstu1}) we obtain
\bea
i{\cal M}^{\mu \nu \rho \sigma}_s \epsilon_\mu(p_1) \epsilon_\nu(p_2) \epsilon_\rho^*(k_1) k_{2 \sigma}&=& -4ig^2\sin(p_1 \theta p_2/2)\sin(k_1 \theta k_2/2)\epsilon_\mu(p_1) \epsilon_\nu(p_2) \epsilon_\rho^*(k_1) \nonumber
 \\ &&\times g_{\alpha \beta}(g^{\rho \beta} -\frac{q_1^\beta q_1^\rho}{q_1^2}  )C^{\alpha \mu \nu}(-p_1,-p_2,q_1),\nonumber \\
&&\nonumber \\
i{\cal M}^{\mu \nu \rho \sigma}_t \epsilon_\mu(p_1) \epsilon_\nu(p_2) \epsilon_\rho^*(k_1) k_{2 \sigma} &=& 4ig^2\sin(p_1 \theta k_1/2)\sin(p_2 \theta k_2/2)\epsilon_\mu(p_1) \epsilon_\nu(p_2) \epsilon_\rho^*(k_1)\nonumber \\
&&\times g_{\alpha \beta}( g^{\nu \beta} -\frac{q_2^\beta q_2^\nu}{q^2_2}  )C^{\mu \alpha \rho}(q_2,k_1,-p_1),\nonumber \\
&&\nonumber \\
i{\cal M}^{\mu \nu \rho \sigma}_u \epsilon_\mu(p_1) \epsilon_\nu(p_2) \epsilon_\rho^*(k_1) k_{2 \sigma}&=& -4ig^2\sin(p_1 \theta k_2/2)\sin(p_2 \theta k_1/2)\epsilon_\mu(p_1) \epsilon_\nu(p_2) \epsilon_\rho^*(k_1) \nonumber \\
&&\times g_{\alpha \beta}(g^{\mu \alpha} -\frac{q_3^\mu q_3^\alpha}{q^2_3}  )C^{\rho \nu \beta}(p_2,q_3,-k_1),
\label{ampqcdstu2}
\eea

The following products vanish:
\bea
&&\epsilon_\mu(p_1) \epsilon_\nu(p_2) \epsilon_\rho^*(k_1)g_{\alpha \beta}q_1^\rho q_1^\beta C^{\alpha \mu \nu}(-p_1,-p_2,q_1) =0,
\nonumber \\
&&\nonumber \\
&&\epsilon_\mu(p_1) \epsilon_\nu(p_2) \epsilon_\rho^*(k_1)g_{\alpha \beta}q_2^\nu q_2^\beta C^{\mu \alpha \rho}(q_2,k_1,-p_1)=0,\nonumber \\
&&\nonumber \\
&&\epsilon_\mu(p_1) \epsilon_\nu(p_2) \epsilon_\rho^*(k_1)g_{\alpha \beta}q_3^\mu q_3^\alpha C^{\rho \nu \beta}(p_2,q_3,-k_1)=0.
\label{rel3}
\eea
With this the amplitudes in (\ref{ampqcdstu2}) take the following expressions   
\bea
&&i{\cal M}^{\mu \nu \rho \sigma}_s \epsilon_\mu(p_1) \epsilon_\nu(p_2) \epsilon_\rho^*(k_1)k_{2\sigma}  =-4ig^2\epsilon_\mu(p_1) \epsilon_\nu(p_2) \epsilon_\rho^*(k_1) \sin(p_1 \theta p_2/2)\sin(k_1 \theta k_2/2)  C^{\rho \mu \nu}(-p_1,-p_2,q_1) ,\nonumber \\
&&\nonumber \\&&i{\cal M}^{\mu \nu \rho \sigma}_t \epsilon_\mu(p_1) \epsilon_\nu(p_2) \epsilon_\rho^*(k_1)k_{2\sigma} =4ig^2\epsilon_\mu(p_1) \epsilon_\nu(p_2) \epsilon_\rho^*(k_1)  \sin(p_1 \theta k_1/2)\sin(p_2 \theta k_2/2) C^{\mu \nu \rho}(q_2,k_1,-p_1),\nonumber \\
&&\nonumber \\
&&i{\cal M}^{\mu \nu \rho \sigma}_u \epsilon_\mu(p_1) \epsilon_\nu(p_2) \epsilon_\rho^*(k_1)k_{2\sigma}  = -4ig^2\epsilon_\mu(p_1) \epsilon_\nu(p_2) \epsilon_\rho^*(k_1)\sin(p_1 \theta k_2/2)\sin(p_2 \theta k_1/2) C^{\rho \nu \mu}(p_2,q_3,-k_1).
\label{ampqcdstu3}
\eea

Now let us consider  the fourth contribution. According to the Feynman rules in FIG. (\ref{fgamma}), the invariant amplitude for such contribution takes the form
\bea
i{\cal M}^{\mu \nu \rho \sigma}_c \epsilon_\mu(p_1) \epsilon_\nu(p_2) \epsilon_\rho^*(k_1)\epsilon_\sigma(k_2) &&= 
4ig^2\epsilon_\mu(p_1) \epsilon_\nu(p_2) \epsilon_\rho^*(k_1)\epsilon_\sigma(k_2)\nonumber \\
&&\times\left[ \sin(p_1 \theta p_2/2)\sin(k_1 \theta k_2/2)
 (g^{\sigma \nu}g^{\mu \rho} -g^{\sigma \mu}g^{\rho \nu})\right. \nonumber \\
&&\left. + \sin(p_1 \theta k_1/2)\sin(p_2 \theta k_2/2)(g^{\sigma \rho}g^{\mu \nu} -g^{\sigma \mu}g^{\nu \rho})\right. \nonumber \\
&&\left. +\sin(p_1 \theta k_2/2)\sin(p_2 \theta k_1/2)( g^{\sigma \rho}g^{\mu \nu}- g^{\sigma \nu}g^{\mu \rho})\right].
\label{termcontact}
\eea
Replacing $\epsilon_\sigma(k_2)$ by $k_{2\sigma}$, we get
\bea
i{\cal M}^{\mu \nu \rho \sigma}_c \epsilon_\mu(p_1) \epsilon_\nu(p_2) \epsilon_\rho^*(k_1)k_{2\sigma} && = 4ig^2\epsilon_\mu(p_1) \epsilon_\nu(p_2) \epsilon_\rho^*(k_1)\nonumber \\
&&\times [\sin(p_1 \theta p_2/2)\sin(k_1 \theta k_2/2)(k_{2\nu}g^{\mu \rho} -k_{2\nu}g^{\rho \nu}) \nonumber \\
&&+ \sin(p_1 \theta k_1/2)\sin(p_2 \theta k_2/2)(k_{2\rho}g^{\mu \nu} - k_{2\mu}g^{\nu \rho})\nonumber \\
&&+\sin(p_1 \theta k_2/2)\sin(p_2 \theta k_1/2)(k_{2\rho}g^{\mu \nu} -k_{2\nu}g^{\mu \rho})].
\label{ampqcdc2}
\eea
Now we have to sum  the four contributions. Using momentum conservation, we can eliminate $q_1$, $q_2$ and $q_3$ in favor of
 $p_1$, $p_2$, $ k_1$ and $k_2$. Then after some manipulation we are able to write the total amplitude in the following simple form  
\bea
&&i{\cal M}^{\mu \nu \rho \sigma}_{total} \epsilon_\mu(p_1) \epsilon_\nu(p_2) \epsilon_\rho^*(k_1)k_{2\sigma}  = -4ig^2\epsilon_\mu(p_1) \epsilon_\nu(p_2) \epsilon_\rho^*(k_1)C^{\rho \mu \nu}(-p_1,-p_2,k_1) \nonumber \\
&&\times\left(\sin(p_1 \theta p_2/2)\sin(k_1 \theta k_2/2)+\sin(k_1 \theta p_1/2)\sin(p_2 \theta k_2/2)
+\sin(p_1 \theta k_2/2)\sin(p_2 \theta k_1/2)\right ).
\label{ampncqedtotal}
\eea

It will be very instructive if we consider in this point of our checking  the Ward identity in the $gg \rightarrow gg$ scattering in QCD. This scattering is very similar to our scattering above both in  number of contributions and in  their Feynman rules. However there is the   subtle difference that  gluons carry color. The Feynman rules used here can be found, for instance, in \cite{peskin}. Making the same 
 assumptions and taking the same steps as done in the case of NCQED above, and also  making
 the same  distribution of momenta and polarization vectors for the four external gluons, we get the following expression for the contraction of the momentum $k_{2 \sigma}$ with the total amplitude of the scattering 
\bea
i{\cal M}^{\mu \nu \rho \sigma}_{QCD} \epsilon_\mu(p_1) \epsilon_\nu(p_2) \epsilon_\rho^*(k_1)k_{2\sigma} &=& -ig^2\epsilon_\nu(p_1) \epsilon_\mu(p_2) \epsilon_\rho^*(k_1)C^{\rho \mu \nu}(-p_1,-p_2,k_1)\nonumber \\
&&\times  ( f^{abc}f^{cfg} + f^{gac}f^{cfb} + f^{fac}f^{cbg}).
\label{ampqcdtotal}
\eea
We can see from (\ref{ampqcdtotal}) that it is the  Jacobi identity
\be
 f^{abc}f^{cfg} + f^{gac}f^{cfb} + f^{fac}f^{cbg}=0, 
\label{jacobi}
\ee
that guarantees  the Ward identity
\be
i{\cal M}^{\mu \nu \rho \sigma}_{QCD} \epsilon_\mu(p_1) \epsilon_\nu(p_2) \epsilon_\rho^*(k_1)k_{2\sigma}=0.
\label{wardgg}
\ee 
This express how important are the structure constants, through the Jacobi identity,  in getting the Ward identity in 
QCD.

 It is interesting to see  if the momentum-dependent phase factors engender an analogous identity. To see that this in fact happens we notice that  the expression in (\ref{ampncqedtotal})  goes to zero if and only if   
\be
\sin(p_1 \theta p_2/2)\sin(k_1 \theta k_2/2)+\sin(k_1 \theta p_1/2)\sin(p_2 \theta k_2/2)
+\sin(p_1 \theta k_2/2)\sin(p_2 \theta k_1/2)=0.
\label{ncidentity}
\ee

Let us suppose  that this is really  true. Then  we can  map  the structure constants of QCD in the moment-dependent phase factors of  NCQED as follows
\bea
&&f^{abc} \leftrightarrow 2\sin(p_1 \theta p_2/2) \,\,\,\,\,\, ,\,\,\,\,f^{cfg}\leftrightarrow 2\sin(k_1 \theta k_2/2),\nonumber \\
&&f^{fac}\leftrightarrow 2\sin(p_1 \theta k_2/2) \,\,\,\,\,\, ,\,\,\,\,f^{cbg}\leftrightarrow 2\sin(p_2 \theta k_1/2),\nonumber \\
&&f^{gac}\leftrightarrow 2\sin(k_1 \theta p_1/2) \,\,\,\,\,\, ,\,\,\,\,f^{cbf}\leftrightarrow 2\sin(k_2 \theta p_2/2).
\label{newrelation}
\eea
The antisymmetry among two indices of the structure constant $f^{abc}$ is equivalent to the antisymmetry among the two  indices of $\theta^{\mu \nu}$ which translates in the property
\be
 \sin(p \theta q/2)=-\sin(q \theta p/2).
\label{antisymmetry}
\ee
For checking that such mapping really works, we notice that in substituting all the momentum-dependent phase factors  by the structure constants in  the calculations of the  $\gamma \gamma \rightarrow \gamma \gamma$ scattering,  we reproduce all the steps of the $gg \rightarrow gg$ scattering in QCD.  The usefulness of this mapping is  that, with some care,  we can obtain the amplitude for any process in NCQED from its similar amplitude in QCD, or vice-versa. For example,    we could soon arrive in (\ref{ampqcdtotal}) from  (\ref{ampncqedtotal}) through the mapping suggested above.

Let us show that in fact  (\ref{ncidentity}) vanishes. For this we make use of the momentum conservation $p_1+p_2 = k_1 + k_2$ for eliminating $k_2$ in favor of the other momenta. Doing this and using (\ref{antisymmetry}), we have that 
\bea
&&\sin(k_1 \theta k_2/2) = \sin(k_1 \theta p_1/2)\cos(p_2 \theta k_1/2)-\cos(p_1 \theta k_1/2)\sin(p_2 \theta k_1/2),\nonumber \\
&&\sin(p_2 \theta   k_2/2) =- \sin(p_1 \theta p_2/2)\cos(p_2 \theta k_1/2)-\cos(p_1 \theta p_2/2)\sin(p_2 \theta k_1/2),\nonumber \\
&&\sin(p_2 \theta  k_1/2) = \sin(p_1 \theta p_2/2)\cos(p_1 \theta k_1/2)+\cos(p_1 \theta p_2/2)\sin(k_1 \theta p_1/2).
\label{opening}
\eea 
Substituting (\ref{opening}) in (\ref{ncidentity}) we easily see that (\ref{ncidentity}) is really true. Then like the structure constants in QCD, the momentum-dependent phase factors in NCQED are crucials in the validity of the Ward identity.

\section{final remarks}
\label{sec5}
In this work we checked the Ward identity in pair annihilation process and   $\gamma \gamma \rightarrow \gamma \gamma$ scattering in the context of NCQED.  As expected, in both processes the Ward identity is satisfied. We emphasize that our check is general, valid for $\theta^{\mu \nu}$ arbitrary.

 In regard to the $\gamma \gamma \rightarrow \gamma \gamma$ scattering, we found a kind of identity among the momentum-dependent phase factors which played a role similar to  the Jacobi identity in QCD. Due to those similarities we have been able to make a mapping among QCD and NCQED. With such mapping we can write all the Feynman rules and also invariant amplitudes in NCQED from  similar processes in QCD. Finaly

\begin{acknowledgments} 
It is a pleasure to thank D. Bazeia for encouragement, and for reading of  the manuscript.  The work of T.M. is supported by Conselho Nacional de Desenvolvimento Cient\'{\i}fico e Tecnol\'ogico(CNPq).
\end{acknowledgments}
\appendix
\section*{Ward identity}
\label{ap}

The Ward identity is a constraint that appears in processes that  present external photons. In those  processes  the invariant amplitude takes the form
\be
{\cal M} = \epsilon_\alpha(k_1)\epsilon_\beta(k_2)\cdots {\cal M}^{\alpha \beta \cdots}(k_1,k_2,\cdots).
\label{amplitude}
\ee
The Ward identity states that for all the external photons we must have 
\be
k_{1\alpha}{\cal M}^{\alpha \beta \cdots}=k_{2\beta}{\cal M}^{\alpha \beta \cdots}= \cdots =0.
\label{geralwardident}
\ee
There are other ways of stating the  Ward identity, as for example  in terms of renormalization factors. Since we restrict our analysis to tree level processes, the above  statement  is more useful for us. In practical terms  (\ref{geralwardident})
demands the on shell photons to  be transverse. For sake of completeness we present a short demonstration of   (\ref{geralwardident}). 

Ward identity is strictly connected to gauge invariance.  In ordinary QED  the gauge field transforms as
\be
A_\mu \rightarrow A_\mu + \partial_\mu \alpha.
\label{gaueinv}
\ee
We can describe  the gauge field, in a Lorentz gauge, by the following plane wave
\be
A_\mu \sim \epsilon_\mu(k)e^{\pm ik\cdot x}.
\label{planewave}
\ee
Taking $\alpha \sim \widetilde{\alpha}(k)e^{\pm ik\cdot x}$ the gauge invariance in (\ref{gaueinv}) translates in the following
 transformation of the polarization vector
\be
\epsilon_\mu(k) \rightarrow \epsilon_\mu(k) \pm k_\mu \widetilde{\alpha}(k).
\label{polariztransf}
\ee
In view of this transformation for the polarization vectors  the invariance of the amplitude ${\cal M}$ in  (\ref{amplitude}) leads to the constraint in  (\ref{geralwardident}). The procedure  in nonabelian
 symmetry  is similar.
%
\def\MPL #1 #2 #3 {Mod. Phys. Lett. A {\bf#1},\ #2 (#3)}
\def\NPB #1 #2 #3 {Nucl. Phys. {\bf#1},\ #2 (#3)}
\def\PLB #1 #2 #3 {Phys. Lett. B {\bf#1},\ #2 (#3)}
\def\PR #1 #2 #3 {Phys. Rep. {\bf#1},\ #2 (#3)}
\def\PRD #1 #2 #3 {Phys. Rev. D {\bf#1},\ #2 (#3)}
\def\PRL #1 #2 #3 {Phys. Rev. Lett. {\bf#1},\ #2 (#3)}
\def\RMP #1 #2 #3 {Rev. Mod. Phys. {\bf#1},\ #2 (#3)}
\def\NIM #1 #2 #3 {Nuc. Inst. Meth. {\bf#1},\ #2 (#3)}
\def\ZPC #1 #2 #3 {Z. Phys. {\bf#1},\ #2 (#3)}
\def\EJPC #1 #2 #3 {E. Phys. J. C {\bf#1},\ #2 (#3)}
\def\IJMP #1 #2 #3 {Int. J. Mod. Phys. A {\bf#1},\ #2 (#3)}
\def\JHEP #1 #2 #3 {J. High Energy Phys. {\bf#1},\ #2 (#3)}


\begin{references}
\bibitem{snyder}
H. S. Snyder,  \PR  71 38 1947 .

\bibitem{connes}
A. Connes, \JHEP 9802 003 1998 .
\bibitem{s-w}
N. Seiberg and E. Witten, \JHEP 9909 032 1999 .

\bibitem{douglas}
 M. R. Douglas and N.  A. Nekrasov, 
\RMP 73 977 2001 .

\bibitem{scalar}
M. V. Raamsdonk and N. Seiberg, \JHEP 03 035 2000 , H. Grosse, C. Klimcik, and P. Presnajder, Commun. Math. Phys., 180, 429 (1996) ,
I. ya. Aref'eva, D. M. Belov and A. S. Koshelev, \PLB 476 431 2000 . 



\bibitem{riad}
Ihab. F. Riad and M.M. Sheikh-Jabbari, \JHEP 08 045 2000 .

\bibitem{hayakama}
M. Hayakama, \PLB 478 394 2000 .

\bibitem{susskind}
A. Matusis, L. Susskind and N. Toumbas,  \JHEP 0012 002 2000 .
\bibitem{lorentz}
S. M. Carroll, J. A. Harvey, V. A. Kostelecky, C.
D. Lane  and  T. Okamoto, \PRL 87 141601 2001 .


\bibitem{unitarity}
 Gomis and T.  Mehen,  \NPB B591 265 2000 ;
C-S Chu, J. Lukierski and W. J. Zakrzewski, \NPB B632 219 2002 ;
Y. Liao and K.  Sibold  Eur. Phys. J. C {\bf25}, 479 (2002). 

\bibitem{frenkel}

F. T. Brandt, J. Frenkel and D. G. C. McKeon, \PRD 65 125029 2002 ,
F. T. Brandt, Ashok Das and J. Frenkel, \PRD 65 085017 2002.

\bibitem{yang-mills}
T. Krajewski and R. Wulkenhaar, \IJMP 15 1011 2000 , A. Armoni, \NPB B593 229 2001 ;
F. T. Brandt, Ashok Das, J. Frenke,  D. G. C.
McKeon and J. C. Taylor, \PRD 66 045011 2002 .

\bibitem{varios} For an incomplete list of works on NCQFT see:
T. Filk, \PLB 376 53 1996 ;
I. Mociou, M. Pospelov and R. Roiban, \PLB  489 390 2000 ;
H. O. Girotti, M. Gomes, V. O. Rivelles and  A. J. da Silva, \NPB B587 299 2000 ;
M. M. Sheikh-Jabbari, \PRL 84 5265 200 ;
 L. Alvarez-Gaum\'e and S. Wadia, \PLB 501 319 2001 ;
 C. P. Martin and F. Ruiz, \NPB B597 197 2001 ;
 C. E. Carlson, C. D. Carone and R.F. Lebed, \PLB 518 201 2001 ; 
A.  Bichl, J.  Grimstrup, H.  Grosse,  L.  Popp, M. Schweda and  R. Wulkenhaar \JHEP 0106 013 2001 ;
 Z. Guralnik, R. Jackiw, S.Y. Pi and A. P. Polychronakos, \PLB 517 450 2001 ;
J. Gamboa, M. Loewe and J. C. Rojas, \PRD 64 067901 2001 ;
 C. E. Carlson,  C. D. Carone and  N. Zobin, \PRD 66 075001-1 2002 ; 
P-M Ho and H-C Kao, \PRL 88 151602-1 2002;
H. Falomir, J. Gamboa, M. Loewe, F. Mendez and J. C. Rojas, \PRD 66 045018-1 2002 ;
 M. Lubo, \PRD 65 066003 2002  ;
 H. O. Girotti, M. Gomes, A. Y. Petrov, V. O. Rivelles and A. J. da Silva,  hep-th/0207220 ; 
 K. Morita,  hep-th/0209234; 
C. E. Carlson and C. D. Carone,  hep-ph/0209077;
C. P. Martin, hep-th/0211164.



\bibitem{review}
For an excelente review on the subject see:
 I. Hinchliffe and N. Kersting, hep-ph/0205040.

\bibitem{mathews}
P. Mathews, \PRD 63 075007-1 2001 ;
T. Rizzo,  hep-ph/0203240;
S. Godfrey and M. A. Doncheski,  hep-ph/0111147;
A. Anisimov, T. Banks, M. Dine, and M. Graesser,
H. Grosse and Y.  Liao  \PRD 64 115007 2001 .


\bibitem{chair}
N. Chair, M. M. Sheikh-Jabbari,  \PLB 504 146 2001 . 

\bibitem{gamma}
J.  L. Hewett, F. J. Petriello and T. G. Rizzo,  \PRD 64 075012 2001 ,  {\it ibid}  \PRD 64 075012 2001 ,
S. Godfrey,and  M. A. Doncheski,  \PRD 65 015005 2002 ,
S-W Baek, D. K. Ghosh, X-G He and  W.Y.P. Hwang, \ PRD 64 056001 2001 .
N. Mahajan,  hep-ph/0110148.

\bibitem{comment}
The works \cite{riad} and \cite{frenkel}  have checked the Ward identity at loop level for two point functions. 


 \bibitem{peskin}
M. E. Peskin and D. V. Schroeder, An Introduction to Quantum Field Theory ( Addison-Wesley, 1995).
\end{references}
\end{document}